\documentclass[11pt]{article}
\usepackage[a4paper]{geometry}
\usepackage[italian,english]{babel}

\usepackage{fancyhdr}
\usepackage{amsmath}
\usepackage{amsfonts}
\usepackage{amstext}
\usepackage{amssymb}
\usepackage{amsthm}
\usepackage{amscd}

\usepackage[pagebackref,draft=false]{hyperref}
\hypersetup{colorlinks,
linkcolor=myrefcolor,
citecolor=mycitecolor,
urlcolor=myurlcolor}

\usepackage[capitalize]{cleveref}
\usepackage{caption}
\usepackage{etaremune}

\usepackage{xcolor}
\definecolor{myurlcolor}{rgb}{0,0,0.4}
\definecolor{mycitecolor}{rgb}{0,0.5,0}
\definecolor{myrefcolor}{rgb}{0.5,0,0}
\usepackage{graphicx}
\usepackage{tikz}
\usepackage{tikz-cd}
\usepackage{mathrsfs}

\usepackage{etoolbox}
\usepackage{makeidx}
\usepackage{sectsty}
\usepackage{dsfont}
\usepackage{enumitem} 
\usepackage[]{latexsym}
\usepackage{braket}
\usepackage{caption}
\usepackage[utf8]{inputenx}
\usepackage[T1]{fontenc}
\usepackage{lmodern}
\usepackage{textcomp}
\usepackage{microtype}
\usepackage{totcount}
\usepackage{blindtext}

\newtheorem{remark}{Remark}

\newtheorem*{proof*}{Proof}

\newcommand{\be}{\begin{equation}}
\newcommand{\ee}{\end{equation}}
\newcommand{\bea}{\begin{eqnarray}}
\newcommand{\eea}{\end{eqnarray}}

\newcommand{\grit}[1]{{\bfseries {\itshape {#1}}}}

\newcommand{\blue}[1]{\textcolor{blue}{{#1}}}

\newcommand{\ra}{\rightarrow}

\newcommand{\lra}{\longrightarrow}

\newcommand{\hh}{\mathcal{H}}
\newcommand{\bh}{\mathcal{B}(\mathcal{H})}
\newcommand{\Glh}{\mathcal{GL}(\mathcal{H})}

\newcommand{\Uh}{\mathcal{U}(\mathcal{H})}
\newcommand{\uh}{\mathfrak{u}(\mathcal{H})}

\newcommand{\Tr}{\textit{Tr}}

\newcommand{\stsp}{\mathscr{S}}

\newcommand{\stav}{\mathscr{V}}

\newcommand{\pos}{\mathscr{P}}

\newcommand{\gr}{\mathrm{g}}

\newcommand{\G}{\mathrm{G}}

\title{Quantum States, Groups and Monotone Metric Tensors}

\author{F. M. Ciaglia$^{1,2}$  \href{https://orcid.org/0000-0002-8987-1181}{\includegraphics[scale=0.7]{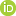}},\\
\footnotesize{$^{1}$\textit{ Max Planck Institute for Mathematics in the Sciences, Leipzig, Germany}} \\
\footnotesize{$^{2}$\textit{ e-mail: \texttt{florio.m.ciaglia[at]gmail.com} and \texttt{ciaglia[at]mis.mpg.de}}}
}

\begin{document}

\maketitle

\begin{abstract}
A novel link between monotone metric tensors and actions of suitable extensions of the unitary group  on the manifold of faithful quantum states is presented here by means of  three illustrative examples related with the Bures-Helstrom metric tensor, the Wigner-Yanase metric tensor, and the Bogoliubov-Kubo-Mori metric tensor.

\end{abstract}


\thispagestyle{fancy}

\section{Introduction}

In this work, a preliminary analysis of the relation between monotone metric tensors on the manifold of faithful quantum states and group actions of  suitable extensions of the unitary group is presented.
In the context of Quantum Information Geometry of finite-dimensional systems with Hilbert space $\hh$, the unitary group $\Uh$ plays the role of universal symmetry group for the class of the so-called {\itshape monotone metric tensors} on the space $\stsp_{+}$ of faithful states (invertible density operators $\hh$) providing the quantum counterpart of the classical Fisher-Rao metric tensor \cite{C-M-1991,Petz-1996}.
Specifically, the unitary group $\Uh$ acts on quantum states (and more generally on the whole space $\stav$ of Hermitean operators on $\hh$) according to the standard action 
\be\label{eqn: action of U(H)}
\phi(\mathbf{U},\,\rho)\,\equiv\,\phi_{\mathbf{U}}(\rho)\,=\,\mathbf{U}\,\rho\,\mathbf{U}^{\dagger},
\ee
where $\rho$ is a quantum state and $\mathbf{U}\in\Uh$, and  $\phi_{\mathbf{U}}$  represents an isometry of every monotone metric tensor $\G$  because of the requirement of monotonicity  under completely-positive, trace-preserving maps representing the quantum version of classical coarse graining \cite{C-M-1991,Petz-1996}.
From the infinitesimal point of view, the action $\phi$ is described in terms of the  fundamental vector fields on $\stsp_{+}$ providing an anti-representation of the Lie algbera $\mathfrak{u}(\mathcal{H})$ of the unitary group.
These vector fields, denoted by $\mathbb{X}_{\mathbf{b}}$ with $\mathbf{b}$ an Hermitian operator on $\hh$ (more on this in section \ref{sec: B-H metric tensor}), are Killing vector fields for \grit{all} the monotone metric tensors  because   $\Uh$ acts by means of isometries.

Now, the Lie algebra $\uh$   is a Lie-subalgebra of the space $\bh$ of bounded linear operators on $\hh$ endowed with the Lie product given by the commutator $[\cdot,\cdot]$ between linear operators.
In particular, it turns out that $\bh$ (endowed with $[\cdot,\cdot]$) is isomorphic to the Lie algebra of the complexification of $\Uh$, namely, to the Lie algebra of the Lie group $\Glh$ consisting of invertible linear operators on $\hh$.
Furthermore, it is known \cite{C-C-I-M-V-2019,C-I-J-M-2019,G-K-M-2005,G-K-M-2006} that $\Glh$ acts on the manifold $\stsp_{+}$, and more generally on the whole space of quantum states $\stsp$,  according to 
\be 
\alpha(\gr,\rho)\,=\,\frac{\gr\,\rho\,\gr^{\dagger}}{\Tr(\gr\rho\gr^{\dagger})},
\ee
where $\gr\in\Glh$.
Clearly, if $\gr\in\Uh\subset\Glh$, we obtain again the standard action $\phi$ of $\Uh$ introduced above.
The fundamental vector fields of $\alpha$ provide an anti-representation of the Lie algebra of $\Glh$ that    ``contains'' the fundamental vector fields $\mathbb{X}_{\mathbf{b}}$ of the action $\phi$ of $\Uh$ together with some {\itshape complementary} vector fields,  written as $\mathbb{Y}_{\mathbf{a}}$ with $\mathbf{a}$ an Hermitean operator on $\hh$ (more on this in section \ref{sec: B-H metric tensor}),  which do not close a Lie algebra on their own.

These complementary vector fields were shown to be essential ingredients in the geometric formulation of the Gorini-Kossakowski-Lindblad-Sudarshan equation governing the dynamical evolution of open quantum systems \cite{C-DC-I-L-M-2017,C-DC-L-M-2017}, while their role in the context of Quantum Information geometry is not completely clear.
As will be noted in section \ref{sec: B-H metric tensor}, the $\mathbb{Y}_{\mathbf{a}}$ are essentially the Symmetrized Logarithmic Derivative introduced by Helstrom  \cite{Helstrom-1967,Helstrom-1968,Helstrom-1969}.
Moreover, it  will also be shown   that the $\mathbb{Y}_{\mathbf{a}}$ are the gradient vector fields associated with the expectation value functions of quantum observables (Hermitean operators on $\hh$) given by
\be
f_{\mathbf{a}}(\rho)\,=\,\Tr(\rho\,\mathbf{a})\,
\ee
by means of  the   monotone metric tensor $\G_{BH}$  known as the Bures-Helstrom metric tensor  \cite{B-Z-2006,Dittmann-1993,Dittmann-1995,Helstrom-1967,Uhlmann-1992}.

Reading this instance backward, we can say that the gradient vector fields associated with the $f_{\mathbf{a}}$ by means of the Bures-Helstrom metric tensor, together with the $\mathbb{X}_{\mathbf{b}}$,  provide a representation of a Lie algebra  ``enlargment'' of $\uh$, namely, the Lie algebra of the complexification $\Glh$ of $\Uh$, and  this representation integrates to the group action $\alpha$ of $\Glh$ described above.

Since the Bures-Helstrom metric tensor $\G_{BH}$ is only one of the (infinitely many) possible quantum counterparts of the Fisher-Rao metric tensor, it is reasonable to ask if $\G_{BH}$ is the only metric tensor among the monotone ones for which such an instance is realized.
Specifically,  given a monotone metric tensor $\G$   (Quantum Fisher-Rao metric tensor), the question is if the gradient vector fields that it is possible to associated with the expectation value functions, together with the universal Killing vector fields $\mathbb{X}_{\mathbf{b}}$, provide a representation of a Lie algebra integrating to a group action.
In the following sections, we will answer   this question  in the affirmative  by providing explicit examples.

It will be proved in section \ref{sec: W-Y metric tensor} that the $\mathbb{X}_{\mathbf{b}}$ together with the gradient vector fields, say $\mathbb{W}_{\mathbf{a}}$, associated with the functions $f_{\mathbf{a}}$ by means of  the monotone metric tensor $\G_{WY}$ known as the Wigner-Yanase metric tensor \cite{G-I-2001,G-I-2003,Hasegawa-2003,H-P-1997}  provide an anti-representation of the Lie algebra of $\Glh$ integrating to the action $\Theta$ of $\Glh$ given by
\be
\Theta (\gr,\,\rho)\,=\frac{\left(\gr\,\sqrt{\rho}\,\gr^{\dagger}\right)^{2}}{\Tr\left(\left(\gr\,\sqrt{\rho}\,\gr^{\dagger}\right)^{2}\right)}.
\ee
This action is clearly different from $\alpha$, but reduces to $\phi$ if unitary elements are considered.
The action $\Theta$ can be extended to the whole space of states $\stsp$, and   it is easy to see that the orbits of this action are precisely the manifolds of quantum states with fixed rank, in analogy with what happens for the action $\alpha$.
In particular, $\Theta$ preserves  the manifold $\stsp_{1}\cong\mathbb{CP}(\hh)$ of pure (rank-1) quantum states, and, since every pure state $\rho_{p}$   satisfies $\rho_{p}^{2}=\rho_{p}$,  it is easy to check that 
\be
\Theta(\gr,\rho_{p})\,=\,\alpha(\gr,\rho_{p})
\ee
for every $\gr\in\Glh$, and every pure state $\rho_{p}$, while, in general, $\alpha$ is different from $\Theta$ on mixed states.

The fact that both actions $\alpha$ and $\Theta$ can be extended to (and agree on) the space $\stsp_{1}\cong\mathbb{CP}(\hh)$ of pure quantum states is particularly interesting  when it is noted that both the Bures-Helstrom metric tensor and the Wigner-Yanase metric tensor belong to the particular subset of monotone metric tensors admitting what is known as a ``radial limit'' to the space of pure quantum states \cite{P-S-1996} which is  (proportional to) the metric tensor on $\stsp_{1}\cong\mathbb{CP}(\hh)$ given by the Fubini-Study metric tensor.
This may suggest that the ``radial limit'' of the monotone metric has something to do with the extension of the associated group action  to the pure quantum states (if the group action exists, of course).

In support to this intuition, it will be shown in section \ref{sec: B-K-M metric tensor} that there is at least one monotone metric tensor which does not admit a ``radial limit'' and whose associated group action does not extend to the space of pure states.
Curiously, this monotone metric tensor is associated with the action  of  an enlargment of $\Uh$ different from $\Glh$, and connected with a structure of Lie algebra on $\bh$ which is different from the standard one given by the commutator.
Specifically, the space $\bh$ may be endowed with a structure of Lie algebra which is isomorphic to the Lie algebra of the cotangent Lie group $T^{*}\Uh\cong\Uh\rtimes_{Ad^{*}}\stav$  of the unitary group \cite{A-G-M-M-1994,A-G-M-M-1998}.
Here, $\stav$ denote the space of Hermitian operators on $\hh$, and $Ad^{*}$ is the co-adjoint action of $\Uh$ on $\stav$ when the latter is idetified with the dual of the Lie algebra $\uh$ (more on this in section \ref{sec: B-K-M metric tensor}).
Up to now, and at the best of the author's knowledge, the role of $T^{*}\Uh$ in Quantum Information Geometry is yet to be properly understood, and, as a preliminary step in this direction, in section \ref{sec: B-K-M metric tensor}, it is proved that  $T^{*}\Uh$ acts on $\stsp_{+}$ by means of the action $\Xi$ given by
\be
\Xi((\mathbf{U},\mathbf{a}),\rho)\, =\,\frac{\mathrm{e}^{\mathbf{U}\ln(\rho)\mathbf{U}^{\dagger} + \mathbf{a}}}{\Tr(\mathrm{e}^{\mathbf{U}\ln(\rho)\mathbf{U}^{\dagger} + \mathbf{a}})}.
\ee
This action reduces to the standard action $\phi$ when restricted to the unitary group $\Uh$.
Moreover, it will be proved that the $\mathbb{X}_{\mathbf{b}}$ together with the gradient vector fields  associated with the $f_{\mathbf{a}}$ by means of the monotone metric tensor $\G_{BKM}$ known as the Bogoliubov-Kubo-Mori metric tensor \cite{F-M-A-2019,Naudts-2018,N-V-W-1975,Petz-1993,P-T-1993}, and denoted by  $\mathbb{Z}_{\mathbf{a}}$, are the fundamental vector fields for the action $\Xi$.
The explicit expression of $\Xi$ given above can not be extended outside the space $\stsp_{+}$ of faithful quantum states because the logarithm of a non-faithful (non-invertible) quantum state is not defined.
One may argue that this is not enough to conclude that $\Xi$ does not extend to the space of pure states because the expression of $\Xi$ given above may suitably change when a non-faithful state is considered.
However, since the tangent vector at $\rho$ associated with the vector field $\mathbb{Z}_{\mathbf{a}}$ is given by  
\be 
\mathbb{Z}_{\mathbf{a}} (\rho)\,=\,\frac{\mathrm{d}}{\mathrm{d}t}\left(\frac{\mathrm{e}^{ \ln(\rho)  + t\mathbf{a}}}{\Tr\left(\mathrm{e}^{ \ln(\rho)  + t\mathbf{a}}\right)}\right)_{t=0}=\int_{0}^{1}\,\mathrm{d}\lambda\,\left(\rho^{\lambda}\,\mathbf{a} \,\rho^{1-\lambda}\right) - \Tr(\rho\,\mathbf{a})\,\rho\,,
\ee
if we take  $\rho$ to be a pure state, then $\rho=|\psi\rangle\langle\psi|$ for some normalized vector in $\hh$, $\rho^{\lambda}=\rho^{1-\lambda}$,  and it is easy to check that $\mathbb{Z}_{\mathbf{a}}(\rho)=\mathbf{0}$.
This means that the infinitesimal version of the action $\Xi$ does not extend properly to the space of pure quantum states.

In section \ref{sec: conclusions}, a brief discussion elucidates the future developments that may be followed starting from the results presented here.
In particular,  it will be argued that the unitary invariance satisfied by every monotone metric tensor $\G$ is powerful enough to impose a particular constraint on the possible Lie algebra enlargment  of $\uh$ which could be linked to $\G$ in the way described above.

Finally, a note on the methodology followed in the rest of the work.
For the sake of simplicity, the case of not-normalized quantum states will be considered first, and the case of normalized states will be discussed after.
This will reduce to the minimum the complexity of the computations involved because the normalized case always requires the introduction of a suitable denominator factor that brings in some additional computational complexity  without, however, changing the conceptual scheme of things.

\section{Bures-Helstrom metric tensor}\label{sec: B-H metric tensor}

Let $\pos_{+}\subset\bh$ be the set of strictly positive operators on $\hh$.
Elements in $\pos_{+}$ can be interpreted as not-normalized quantum states, and they form an homogeneus space of the general linear group $\mathcal{GL}(\hh)$ with respect to the action \cite{C-I-J-M-2019,G-K-M-2005,G-K-M-2006}
\be\label{eqn: action og GL(H)}
\widetilde{\alpha}(\gr,\,\rho)\,:=\,\gr\,\rho\,\gr^{\dagger}\,.
\ee
The Lie algebra of $\mathcal{GL}(\hh)$ is the whole algebra $\bh$ of bounded linear operators on $\hh$ endowed with the canonical commutator operator.
Since every element in $\bh$ can be uniquely written as the sum of an Hermitian element $\mathbf{a}$ and a skew-Hermitian element $i\mathbf{b}$, where $\mathbf{b}$ is self-adjoint, the fundamental vector fields  of $\alpha$ are indexed by couples $(\mathbf{a},\mathbf{b})$.
These vector fields are written as $\widetilde{\Gamma}_{\mathbf{ab}}$.
Considering the smooth curve
\be
\gr_{t}\,=\,\mathrm{e}^{\frac{t}{2}(\mathbf{a} + i\mathbf{b})},
\ee
the tangent vector $\widetilde{\Gamma}_{\mathbf{ab}}(\rho)$ at $\rho$ can easily be computed to be
\be
\begin{split}
\widetilde{\Gamma}_{\mathbf{ab}}(\rho)&\,=\,\frac{\mathrm{d}}{\mathrm{d}t}\,\left(\widetilde{\alpha}(\gr_{t},\rho)\right)_{t=0}\,=\,\frac{1}{2}\,\left(\rho\,\mathbf{a} + \mathbf{a}\,\rho \right) + \frac{1}{2 i}\,\left(\rho\,\mathbf{b} - \mathbf{b}\,\rho\right)\,.
\end{split}
\ee
We may decompose $\Gamma_{\mathbf{ab}}$ according to
\be
\Gamma_{\mathbf{ab}}\,=\,\mathcal{Y}_{\mathbf{a}} + \mathbb{X}_{\mathbf{b}},
\ee
where the vector fields $\mathcal{Y}_{\mathbf{a}}$ and $\mathbb{X}_{\mathbf{b}}$ are such that
\be
\begin{split}
\mathcal{Y}_{\mathbf{a}}(\rho)&\,:=\, \frac{1}{2}\,\left(\rho\,\mathbf{a} + \mathbf{a}\,\rho \right) \,\equiv\,\{\rho,\,\mathbf{a}\} \\ 
\mathbb{X}_{\mathbf{b}}(\rho)&\,:=\,\frac{1}{2 i}\,\left(\rho\,\mathbf{b} - \mathbf{b}\,\rho\right)\,\equiv\,[[\rho,\,\mathbf{b}]]\,.
\end{split}
\ee
The vector fields $\mathbb{X}_{\mathbf{b}}$ provide an anti-representation of the Lie algebra $\uh$ of the unitary group $\Uh$ \cite{C-I-J-M-2019,G-K-M-2005,G-K-M-2006} integrating to the action $\phi$ of $\Uh$ mentioned in the introduction and given by equation \eqref{eqn: action of U(H)}.
On the other hand, the vector fields $\mathcal{Y}_{\mathbf{a}}$ do not close a Lie algebra, and may be thought of as complementary vector fields that are needed in order to enlarge the anti-representation of $\uh$  to an anti-representation of the Lie algebra of the complexification $\mathcal{GL}(\hh)$ of $\Uh$.
Moreover, it is worth noting that the tangent vector $\mathcal{Y}_{\mathbf{a}}(\rho)$  provides a geometrical version of the Symmetric Logarithmic Derivative at $\rho$ widely used in quantum estimation theory and metrology  \cite{Helstrom-1967,Helstrom-1968,Helstrom-1969,Paris-2009,Suzuki-2019,T-A-2014}.

It will now be proved that the vector field $\mathcal{Y}_{\mathbf{a}}$ is the gradient vector field associated with the function $f_{\mathbf{a}}$ given by
\be
f_{\mathbf{a}}(\rho)\,=\,\mathrm{Tr}(\rho\,\mathbf{a})
\ee
by means of the Bures-Helstrom metric tensor $\G_{BH}$ on $\pos_{+}$.
The Bures-Helstrom metric tensor $\G_{BH}$ \cite{B-Z-2006,Dittmann-1993,Dittmann-1995,Helstrom-1967,Uhlmann-1992} may be extracted from the Bures distance
\be\label{eqn: Bures distance}
\mathrm{d}_{B}^{2}(\rho,\,\sigma)\,=\,2\,\left(\Tr\rho + \Tr\sigma- 2\Tr\left(\sqrt{\sqrt{\rho}\,\sigma\,\sqrt{\rho}}\right) \right)
\ee
according to the general procedure used in (Classical and Quantum) Information Geometry and given by
\be\label{eqn: from divergence to metric tensor}
 \left(\G_{BH}(X,Y)\right)(\rho)=-\frac{\mathrm{d}}{\mathrm{d}t}\frac{\mathrm{d}}{\mathrm{d}s} \left(\mathrm{d}_{B}^{2}\left(\gamma^{X}_{\rho}(t),\gamma^{Y}_{\rho}(s)\right)\right)_{t,s=0},
\ee
where $X$ and $Y$ are arbitrary vector fields, and $\gamma^{X}_{\rho}(t)$ and $\gamma^{Y}_{\rho}(s)$ their integral curves starting at $\rho$.

\begin{remark}
Note that, in the literature, the Bures-Helstrom metric tensor $\G_{BH}$ is also called ``Quantum Fisher Metric'' \cite{B-C-1994,L-Y-L-W-2020,S-A-G-P-2020,Safranek-2017,Safranek-2018}, while the Bures  metric tensor $\G_{B}$ is usually defined to be the metric tensor associated with half the distance function in equation \eqref{eqn: Bures distance}, and it holds $\G_{BH}=4\G_{B}$. 
Moreover, what will be proved below for the Bures-Helstrom metric tensor $\G_{BH}$ holds also for the Bures metric tensor $\G_{B}$ provided we replace $\mathrm{d}_{B}^{2}$ with  $\frac{1}{2}\mathrm{d}_{B}^{2}$, and   $\mathbf{a}$ with $2\mathbf{a}$ in all the expressions.
The choice made here of using $\G_{BH}$ instead of $\G_{B}$ is essentially due to the fact that $\G_{BH}$ may be thought of as the ``natural'' counterpart of the Fisher-Rao metric tensor in the context of (finite-dimensional) von Neumann algebras as explained in \cite{C-J-S-2020}.
\end{remark}

To actually prove that $\mathcal{Y}_{\mathbf{a}}$ is the gradient vector field associated with $f_{\mathbf{a}}$ by means of $\G_{BH}$, we have to prove that 
\be
\G_{BH}(X,\,\mathcal{Y}_{\mathbf{a}})\,=\,\mathcal{L}_{X}f_{\mathbf{a}},
\ee
where $X$ is an arbitrary vector field and $\mathcal{L}_{X}$ denote the Lie derivative with respect to $X$.
We start computing 
\be
 \left(\G_{BH}(X,\,\mathcal{Y}_{\mathbf{a}})\right)(\rho)=4\frac{\mathrm{d}}{\mathrm{d}t}\frac{\mathrm{d}}{\mathrm{d}s} \Tr\left(\sqrt{\sqrt{\gamma^{X}_{\rho}(t)}\gamma_{\rho}^{\mathcal{Y}_{\mathbf{a}}}(s)\sqrt{\gamma_{\rho}^{X}(t)}} \right)_{s,t=0}.
\ee
To perform the computation, we define the strictly positive operator $\mathbf{C}_{s,t}$, depending parametrically on $s$ and $t$, given by
\be
\mathbf{C}_{s,t}\,:=\,\sqrt{ \gamma_{\rho}^{X}(t)}\gamma_{\rho}^{\mathcal{Y}_{\mathbf{a}}}(s)\sqrt{ \gamma_{\rho}^{X}(t)}.
\ee
Then, we perform a series expansion for $\sqrt{\mathbf{C}_{s,t}} $ around the identity matrix $\mathbb{I}$  obtaining
\be
\begin{split}
\sqrt{\mathbf{C}_{s,t}}&=\sum_{k=0}^{\infty}c_{k}\left(\mathbf{C}_{s,t} -  \mathbb{I} \right)^{k} \\
& =\sum_{k=0}^{\infty}c_{k}\left(\sqrt{ \gamma_{\rho}^{X}(t)}\gamma_{\rho}^{\mathcal{Y}_{\mathbf{a}}}(s)\sqrt{ \gamma_{\rho}^{X}(t)} -  \mathbb{I} \right)^{k}. 
\end{split}
\ee
Using the Leibniz rule and the cyclicity of the trace,  the expression for the series expansion of the derivative of a function, and the Leibniz rule again, we obtain
\be\label{eqn: BH metric tensor}
\begin{split}
\left(\G_{BH}(X,\,\mathcal{Y}_{\mathbf{a}})\right)(\rho)&=4\frac{\mathrm{d}}{\mathrm{d}t} \frac{\mathrm{d}}{\mathrm{d}s}  \sum_{k=0}^{\infty}\,c_{k}\,\left.\Tr \left(\mathbf{C}_{s,t} -  \mathbb{I} \right)^{k}\right|_{t,s=0} \\
&=4\frac{\mathrm{d}}{\mathrm{d}s} \left(\sum_{k=1}^{\infty} c_{k}\,k\Tr \left(\left(\mathbf{C}_{s,0}\right)^{k-1}\frac{\mathrm{d}}{\mathrm{d}t}\left( \mathbf{C}_{s,t}\right)_{t=0} \right) \right)_{s=0} \\
&= 2\frac{\mathrm{d}}{\mathrm{d}s} \left(\Tr \left(\left(\mathbf{C}_{s,0}\right)^{-\frac{1}{2}}\,\frac{\mathrm{d}}{\mathrm{d}t}\left( \mathbf{C}_{s,t}\right)_{t=0} \right) \right)_{s=0} \\
&= 2\Tr \left(\rho^{-1}\,\frac{\mathrm{d}}{\mathrm{d}s}\frac{\mathrm{d}}{\mathrm{d}t}\left( \mathbf{C}_{s,t}\right)_{s,t=0} \right) +  2\Tr \left(\frac{\mathrm{d}}{\mathrm{d}s}\left(\left(\mathbf{C}_{s,0}\right)^{-\frac{1}{2}} \right)_{s=0}\,\frac{\mathrm{d}}{\mathrm{d}t}\left( \mathbf{C}_{0,t}\right)_{t=0} \right)
\end{split}
\ee
Now, we    recall the equality \cite{Suzuki-1997}
\be\label{eqn: derivative of square root}
\frac{\mathrm{d}}{\mathrm{d}t}\left(\sqrt{ \gamma_{\rho}^{X}(t)}\right)_{t=0}\,=\,\mathcal{A}_{\sqrt{\rho}}^{-1}(X(\rho)), 
\ee
where we introduced the superoperator $\mathcal{A}_{\sqrt{\rho}}(\mathbf{b})\,=\,\sqrt{\rho}\,\mathbf{b} + \mathbf{b}\sqrt{\rho}$.
Accordingly, using the Leibniz rule and equation \eqref{eqn: derivative of square root}, we have
\be
\begin{split}
\frac{\mathrm{d}}{\mathrm{d}t}\left( \mathbf{C}_{s,t}\right)_{t=0}&=\frac{\mathrm{d}}{\mathrm{d}t}\left( \sqrt{ \gamma_{\rho}^{X}(t)}\gamma_{\rho}^{\mathcal{Y}_{\mathbf{a}}}(s)\sqrt{ \gamma_{\rho}^{X}(t)}\right)_{t=0} \\
&=\mathcal{A}_{\sqrt{\rho}}^{-1}(X(\rho))\gamma_{\rho}^{\mathcal{Y}_{\mathbf{a}}}(s)\sqrt{\rho} +  \sqrt{\rho} \gamma_{\rho}^{\mathcal{Y}_{\mathbf{a}}}(s)\mathcal{A}_{\sqrt{\rho}}^{-1}(X(\rho)),
\end{split}
\ee
which implies
\be\label{eqn: BH metric tensor 3}
\frac{\mathrm{d}}{\mathrm{d}t}\left( \mathbf{C}_{0,t}\right)_{t=0}=\mathcal{A}_{\sqrt{\rho}}^{-1}(X(\rho)) \rho^{\frac{3}{2}} + \rho^{\frac{3}{2}}\mathcal{A}_{\sqrt{\rho}}^{-1}(X(\rho)),
\ee
and   we also have
\be\label{eqn: BH metric tensor 2}
\begin{split}
\frac{\mathrm{d}}{\mathrm{d}s}\frac{\mathrm{d}}{\mathrm{d}t}\left( \mathbf{C}_{s,t}\right)_{s,t=0}&= \mathcal{A}_{\sqrt{\rho}}^{-1}(X(\rho)) \mathcal{Y}_{\mathbf{a}}(\rho)\sqrt{\rho}  +  \sqrt{\rho}  \mathcal{Y}_{\mathbf{a}}(\rho)\mathcal{A}_{\sqrt{\rho}}^{-1}(X(\rho)).
\end{split}
\ee

Inserting equation \eqref{eqn: BH metric tensor 2}   into equation \eqref{eqn: BH metric tensor}, recalling that 
\be\label{eqn: BH gradient vector fields}
\mathcal{Y}_{\mathbf{a}}(\rho)=\{\rho,\,\mathbf{a}\}=\frac{1}{2}\left(\rho\mathbf{a} + \mathbf{a}\rho\right),
\ee
and exploiting again the cyclicity of the trace, we obtain
\be\label{eqn: BH metric tensor 4}
\begin{split}
\left(\G_{BH}(X,\,\mathcal{Y}_{\mathbf{a}})\right)(\rho)&= \Tr\left(\mathbf{a}\,X(\rho)\right) +   \Tr\left(\rho^{-\frac{1}{2}}\left(\mathcal{A}_{\sqrt{\rho}}^{-1}(X(\rho))\rho\mathbf{a} + \mathbf{a}\rho\mathcal{A}_{\sqrt{\rho}}^{-1}(X(\rho))\right)\right) + \\ 
& \quad +  2\Tr \left(\frac{\mathrm{d}}{\mathrm{d}s}\left(\left(\mathbf{C}_{s,0}\right)^{-\frac{1}{2}} \right)_{s=0}\,\frac{\mathrm{d}}{\mathrm{d}t}\left( \mathbf{C}_{0,t}\right)_{t=0} \right).
\end{split}
\ee

To perform the last derivative with respect to $s$, we exploit the identity \cite{Suzuki-1997}
\be 
\frac{\mathrm{d}}{\mathrm{d}s}\left((\sqrt{\rho}\,\gamma_{\rho}^{\mathcal{Y}_{\mathbf{a}}}(s)\,\sqrt{\rho})^{-\frac{1}{2}}\right)_{s=0}\,=\,- \rho^{-\frac{1}{2}}\, \mathcal{A}_{\rho}^{-1}(\mathcal{Y}_{\mathbf{a}}(\rho))\,\rho^{-\frac{1}{2}},
\ee
which becomes
\be\label{eqn: derivative of -1/2}
\frac{\mathrm{d}}{\mathrm{d}s}\left((\sqrt{\rho}\,\gamma_{\rho}^{\mathcal{Y}_{\mathbf{a}}}(s)\,\sqrt{\rho})^{-\frac{1}{2}}\right)_{s=0}\,=\,- \frac{1}{2}\rho^{-\frac{1}{2}}\,  \mathbf{a} \,\rho^{-\frac{1}{2}}
\ee
because of equation \eqref{eqn: BH gradient vector fields}.
Inserting equation \eqref{eqn: derivative of -1/2} and equation \eqref{eqn: BH metric tensor 3} into equation \eqref{eqn: BH metric tensor 4}, and exploiting once again the ciclicity of the trace, we arrive at the final  expression
\be\label{eqn: BH metric tensor 5}
\begin{split}
 \left(\G_{BH}(X,\,\mathcal{Y}_{\mathbf{a}})\right)(\rho)= \Tr\left(\mathbf{a}\,X(\rho)\right) 
\end{split}
\ee
On the other hand, the Lie derivative $\mathcal{L}_{X}f_{\mathbf{a}}$ is easily seen to be
\be\label{eqn: lie derivative of qrv}
\begin{split}
\left(\mathcal{L}_{X}f_{\mathbf{a}}\right)(\rho)& = \frac{\mathrm{d}}{\mathrm{d}t}f_{\mathbf{a}}\left(\gamma_{\rho}^{X}(t)\right)_{t=0}\\
&=\frac{\mathrm{d}}{\mathrm{d}t}\Tr\left(\mathbf{a}\,\gamma_{\rho}^{X}(t)\right)_{t=0} \\
&= \Tr\,( \mathbf{a} \,X(\rho)) ,
\end{split}
\ee
so that  we have
\be
\G_{ BH }\left(X,\,\mathcal{Y}_{\mathbf{a}}\right)\,=\,\mathcal{L}_{X}f_{\mathbf{a}}, 
\ee
and thus $\mathcal{Y}_{\mathbf{a}}$ is actually the gradient vector field associated with $f_{\mathbf{a}}$ by means of the Bures-Helstrom metric tensor $\G_{BH}$.

\subsection*{Normalized states}

The normalized case is easily obtained from the not-normalized one.
Indeed, let $\stsp_{+}$ denote the manifold of faithful, normalized quantum states.
This is the submanifold of $\pos_{+}$ determined by the (affine) constraint $\Tr\rho=1$ for every $\rho\in\stsp_{+}$.
It is easily seen that the action $\alpha$ given in equation \eqref{eqn: action og GL(H)} does not preserve $\stsp_{+}$ unless we restrict to the unitary group, which means that the $\mathbb{X}_{\mathbf{b}}$'s are tangent to $\stsp_{+}$.
However, it is possible to deform the action $\widetilde{\alpha}$ to an action $\alpha$  preserving $\stsp_{+}$ by setting  \cite{C-C-I-M-V-2019,C-DC-I-L-M-2017,C-I-J-M-2019,G-K-M-2006}
\be
\alpha(\gr,\rho)\,:=\,\frac{\widetilde{\alpha}(\gr,\rho)}{\Tr(\widetilde{\alpha}(\gr,\rho))}\,=\,\frac{\gr\,\rho\,\gr^{\dagger}}{\Tr(\gr\,\rho\,\gr^{\dagger})}.
\ee
Then, we may proceed following the steps outlined before, thus obtaining the fundamental vector fields
\be
\Gamma_{\mathbf{ab}} \,=\,\mathbb{Y}_{\mathbf{a}} + \mathbb{X}_{\mathbf{b}},
\ee
where the $\mathbb{X}_{\mathbf{b}}$'s are the vector fields generating the standard action of the unitary group, and the $\mathbb{Y}_{\mathbf{a}}$'s are given by
\be 
\mathbb{Y}_{\mathbf{a}}(\rho)\,=\,\frac{\mathrm{d}}{\mathrm{d}t}\left(\frac{\mathrm{e}^{\frac{t}{2}\mathbf{a}}\, \rho \,\mathrm{e}^{\frac{t}{2}\mathbf{a}}}{\Tr\left(\mathrm{e}^{\frac{t}{2}\mathbf{a}}\, \rho \,\mathrm{e}^{\frac{t}{2}\mathbf{a}}\right)} \right)_{t=0}\,=\,\{\rho,\,\mathbf{a}\} - \Tr(\rho\mathbf{a})\,\rho.
\ee

The Bures-Helstrom metric tensor on $\stsp_{+}$ is the metric tensor associated with the pullback to $\stsp_{+}$ (with respect to the canonical immersion) of the Bures distance in equation \eqref{eqn: Bures distance}.
With an evident abuse of notation, we denote by $\mathrm{d}_{B}^{2}$  the pullback to $\stsp_{+}$ of the Bures distance given by
\be
\mathrm{d}_{B}^{2}(\rho,\,\sigma)\,=\, 4\left(1- \Tr\left(\sqrt{\sqrt{\rho}\,\sigma\,\sqrt{\rho}}\right)\right)  
\ee
and by $\G_{BH}$ the associated metric tensor on $\stsp_{+}$.

Then, the computations performed in the case of not-normalized states may be easly adapted to prove that
\be\label{eq: gradient vector field for BH}
\G_{BH}(X,\mathbb{Y}_{\mathbf{a}} )\,=\,\mathcal{L}_{X}f_{\mathbf{a}} \,,
\ee
for every vector field $X$ on $\stsp_{+}$, and where $f_{\mathbf{a}} $ is the pullback to $\stsp_{+}$ of the smooth function $f_{\mathbf{a}}$ on $\pos_{+}$, with another evident abuse of notation.
From this, we conclude that every vector field $\mathbb{Y}_{\mathbf{a}}$ is the gradient vector field associated with the smooth function $f_{\mathbf{a}}$ by means of the Bures-Helstrom metric tensor $\G_{BH}$.

\section{Wigner-Yanase  metric tensor}\label{sec: W-Y metric tensor}

Let us consider the diffeomorphism $\varphi\colon\pos_{+}\lra\pos_{+}$ given by
\be\label{eqn: square root diffeo}
\varphi(\rho)\,:=\,\sqrt{\rho}, 
\ee
and its inverse
\be
\varphi^{-1}(\rho)\,=\,\rho^{2}.
\ee
By means of this diffeomorphism, we may define another  action $\widetilde{\Theta}$ of $\Glh$ on $\pos_{+}$ given by
\be
\widetilde{\Theta}(\gr,\,\rho)\,:=\,\varphi^{-1}\,\circ\,\widetilde{\alpha}_{\gr}\,\circ\,\varphi(\rho)\,=\,\left(\gr\,\sqrt{\rho}\,\gr^{\dagger}\right)^{2},
\ee
where we have set $\widetilde{\alpha}_{\gr}(\rho)=\widetilde{\alpha}(\gr,\rho)$ with $\widetilde{\alpha}$ the action given in equation \eqref{eqn: action og GL(H)}.
Clearly, $\widetilde{\Theta}$ is different from $\widetilde{\alpha}$ in general, but, if we take $\gr=\mathbf{U}$ in the unitary group, we have
\be
\widetilde{\Theta}(\mathbf{U},\,\rho)\,=\,\left(\mathbf{U}\,\sqrt{\rho}\,\mathbf{U}^{\dagger}\right)^{2}\,=\,\mathbf{U}\,\rho\,\mathbf{U}^{\dagger}\,=\,\phi(\mathbf{U},\,\rho).
\ee
Following the steps outlined in the previous section, it is easy to show that the fundamental vector fields  $\widetilde{\Psi}_{\mathbf{ab}}$ of $\widetilde{\Theta}$ decompose as
\be
\widetilde{\Psi}_{\mathbf{ab}}\,=\,\mathcal{W}_{\mathbf{a}} + \mathbb{X}_{\mathbf{b}}
\ee
where the $\mathbb{X}_{\mathbf{b}}$'s are the vector fields generating the standard action of the unitary group, and the $\mathcal{W}_{\mathbf{a}}$'s are given by
\be\label{eqn: W vector fields}
\mathcal{W}_{\mathbf{a}}(\rho)\,=\,\frac{\mathrm{d}}{\mathrm{d}t}\left(\left(\mathrm{e}^{\frac{t}{2}\mathbf{a}}\,\sqrt{\rho}\,\mathrm{e}^{\frac{t}{2}\mathbf{a}}\right)^{2}\right)_{t=0}\,=\,\{\rho,\,\mathbf{a}\} + \sqrt{\rho}\,\mathbf{a}\,\sqrt{\rho}.
\ee
Furthermore, note that $\mathcal{W}_{\mathbf{a}}=\varphi_{*}^{-1}\mathcal{Y}_{\mathbf{a}}$, that is, $\mathcal{W}_{\mathbf{a}}$ is the pushforward of $\mathcal{Y}_{\mathbf{a}}$ by means of $\varphi^{-1}$   (see proposition 4.2.4 in \cite{A-M-R-1988}).

We will now prove that every $\mathcal{W}_{\mathbf{a}}$ is the gradient vector field associated with the smooth function $f_{\mathbf{a}}$ by means of the Wigner-Yanase metric tensor $\G_{WY}$ \cite{G-I-2001,G-I-2003,Hasegawa-2003,H-P-1997}.
The Wigner-Yanase metric tensor $\G_{WY}$  is the metric tensor associated with the Wigner-Yanase skew information  
\be\label{eqn: Wigner-Yanase skew information}
S_{WY}(\rho,\,\sigma)\,=\,\, \Tr\rho + \Tr\sigma - 2\Tr\left(\sqrt{\rho}\,\sqrt{\sigma}\right) .
\ee

\begin{remark}
Note that, in the literature, the Wigner-Yanase metric tensor $\G_{WY}$ is usually defined as the metric tensor associated with twice the divergence function in equation \eqref{eqn: Wigner-Yanase skew information}, and  what will be proved below can be easily adapted provided we replace $S_{WY}$ with  $2S_{WY}$, and   $\mathbf{a}$ with $\frac{1}{2}\mathbf{a}$ in all the expressions.
The choice made here  leads to the fact that  the vector fields   $\mathcal{W}_{\mathbf{a}}$ are  the pushforward of the vector fields $\mathcal{Y}_{\mathbf{a}}$ by means of $\varphi^{-1}$   (and similarly for the normalized ones in the following subsection).
\end{remark}
 
Just as we did for the Bures-Helstrom metric tensor in section \ref{sec: B-H metric tensor}, we have to show that  
\be
\G_{ WY }\left(X,\,\mathcal{W}_{\mathbf{a}}\right)\,=\,\mathcal{L}_{X}f_{\mathbf{a}},
\ee
where $X$ is an arbitrary vector field and $\mathcal{L}_{X}$ denote the Lie derivative with respect to $X$.
Therefore, we must compute
\be
\begin{split}
\left(\G_{ WY }\left(X,\,\mathcal{W}_{\mathbf{a}}\right)\right)(\rho)  & = -\frac{\mathrm{d}}{\mathrm{d}t}\,\frac{\mathrm{d}}{\mathrm{d}s}\,\left(D_{WY}(\gamma_{\rho}^{X}(t),\,\gamma_{\rho}^{\mathcal{W}_{\mathbf{a}}}(s))\right)_{t,s=0} \\
& \\
& = 2\frac{\mathrm{d}}{\mathrm{d}t}\frac{\mathrm{d}}{\mathrm{d}s}\left(   \mathrm{Tr} \sqrt{ \gamma_{\rho}^{X}(t)}\,\sqrt{\gamma_{\rho}^{\mathcal{W}_{\mathbf{a}}}(s)} \right)_{t,s=0}
\end{split}
\ee
From equation \eqref{eqn: W vector fields}, we immediately conclude that
\be
\sqrt{\gamma_{\rho}^{\mathcal{W}_{\mathbf{a}}}(s)}\,=\, \mathrm{e}^{\frac{s}{2}\mathbf{a}}\,\sqrt{\rho}\,\mathrm{e}^{\frac{s}{2}\mathbf{a}}\,,
\ee
and recalling equation \eqref{eqn: derivative of square root} we obtain
\be 
\frac{\mathrm{d}}{\mathrm{d}t}\left(\sqrt{ \gamma_{\rho}^{X}(t)}\right)_{t=0}\,=\,\mathcal{A}_{\sqrt{\rho}}^{-1}(X(\rho)), 
\ee
where we introduced the superoperator $\mathcal{A}_{\sqrt{\rho}}(\mathbf{b})\,=\,\sqrt{\rho}\,\mathbf{b} + \mathbf{b}\sqrt{\rho}$.
Consequently, we get
\be
\begin{split}
\left(\G_{ WY }\left(X,\mathcal{W}_{\mathbf{a}}\right)\right)(\rho) & = 2\frac{\mathrm{d}}{\mathrm{d}t}\frac{\mathrm{d}}{\mathrm{d}s}\left(  \mathrm{Tr} \sqrt{ \gamma_{\rho}^{X}(t)}\sqrt{\gamma_{\rho}^{\mathcal{W}_{\mathbf{a}}}(s)} \right)_{s,t=0} \\
&=2 \frac{\mathrm{d}}{\mathrm{d}s}\left( \Tr\left( \mathcal{A}_{\sqrt{\rho}}^{-1}(X(\rho))\sqrt{\gamma_{\rho}^{\mathcal{W}_{\mathbf{a}}}(s)}  \right)\right)_{s =0} \\
&=\Tr\left(\mathcal{A}_{\sqrt{\rho}}^{-1}(X(\rho))\left(\mathbf{a}\,\sqrt{\rho}+ \sqrt{\rho}\,\mathbf{a}\right)\right) \\
&=\Tr\,\left(\mathbf{a}\,\mathcal{A}_{\sqrt{\rho}}\left(\mathcal{A}_{\sqrt{\rho}}^{-1}(X(\rho))\right)\right) \\
&= \Tr(\mathbf{a}\,X(\rho)) .
\end{split}
\ee
Therefore,  recalling equation \eqref{eqn: lie derivative of qrv}, we conclude that 
\be
\G_{ WY }\left(X,\,\mathbb{W}_{\mathbf{a}}\right)\,=\,\mathcal{L}_{X}f_{\mathbf{a}}
\ee
holds for all vector fields $X$ as desired.

\subsection*{Normalized states}

To tackle the normalized case, we mimick what has been done for the Bures-Helstrom metric tensor in section \ref{sec: B-H metric tensor}.
Specifically, we define a normalized version of the action  $\widetilde{\Theta}$, given by
\be
\Theta (\gr,\,\rho)\,:=\,\frac{\widetilde{\Theta}(\gr,\,\rho)}{\Tr(\widetilde{\Theta}(\gr,\,\rho))}=\frac{\left(\gr\,\sqrt{\rho}\,\gr^{\dagger}\right)^{2}}{\Tr\left(\left(\gr\,\sqrt{\rho}\,\gr^{\dagger}\right)^{2}\right)}.
\ee
Then, the fundamental vector fields  $\Psi_{\mathbf{ab}} $ of $\Theta $ are easily computed to be 
\be
\Psi_{\mathbf{ab}} \,=\,\mathbb{W}_{\mathbf{a}}  + \mathbb{X}_{\mathbf{b}}
\ee
where the $\mathbb{X}_{\mathbf{b}}$'s are the vector fields generating the standard action of the unitary group, and the $\mathbb{W}_{\mathbf{a}} $'s are given by
\be\label{eqn: W-normalized vector fields}
\mathbb{W}_{\mathbf{a}} (\rho)\,=\,\frac{\mathrm{d}}{\mathrm{d}t}\left(\frac{\left(\mathrm{e}^{\frac{t}{2}\mathbf{a}}\,\sqrt{\rho}\,\mathrm{e}^{\frac{t}{2}\mathbf{a}}\right)^{2}}{\Tr\left(\left(\mathrm{e}^{\frac{t}{2}\mathbf{a}}\,\sqrt{\rho}\,\mathrm{e}^{\frac{t}{2}\mathbf{a}}\right)^{2}\right)}\right)_{t=0}= \{\rho,\,\mathbf{a}\} + \sqrt{\rho}\,\mathbf{a}\,\sqrt{\rho} - 2\Tr(\rho\mathbf{a})\,\rho.
\ee

Then, the Wigner-Yanase metric tensor on $\stsp_{+}$ is associated with the pullback  to $\stsp_{+}$ (with respect to the canonical immersion) of the Wigner-Yanase skew information given in equation \eqref{eqn: Wigner-Yanase skew information}.
With an evident abuse of notation, we denote by $S_{WY}$  the pullback to $\stsp_{+}$ of the Wigner-Yanase skew information given by
\be
S_{WY}(\rho,\,\sigma)\,=\, 2\left(1- \Tr\left( \sqrt{\rho}\,\sqrt{\sigma}\right)\right)  
\ee
and by $\G_{WY}$ the associated metric tensor on $\stsp_{+}$.

Just as for the Bures-Helstrom metric tensor, the computations performed in the not-normalized case may be easly adapted to prove that
\be\label{eq: gradient vector field for WY}
\G_{WY} (X,\mathbb{W}_{\mathbf{a}} )\,=\,\mathcal{L}_{X}f_{\mathbf{a}} \,,
\ee
for every vector field $X$ on $\stsp_{+}$, and where, with an evident abuse of notation, $f_{\mathbf{a}} $ is the pullback to $\stsp_{+}$ of the smooth function $f_{\mathbf{a}}$ on $\pos_{+}$.
From this, we conclude that every vector field $\mathbb{W}_{\mathbf{a}} $ is the gradient vector field associated with the smooth function $f_{\mathbf{a}} $ by means of the Wigner-Yanase metric tensor $\G_{WY}$.

\section{Bogoliubov-Kubo-Mori metric tensor}\label{sec: B-K-M metric tensor}

The manifold $\pos_{+}$ is an open subset of the vector space $\stav $ of Hermitean (self-adjoint) linear operators on $\hh$.
Moreover, every $\mathbf{h}\in\stav$ gives rise to an element in $\pos_{+}$ by means of  $\mathrm{e}^{\mathbf{h}}$, and every $\rho\in\pos_{+}$ gives rise to an element in $\stav$ by means of $\ln(\rho)$.
Essentially, the map $\psi\colon\pos_{+}\ra\stav$ given by
\be
\psi(\rho)\,:=\,\ln(\rho)
\ee
is a diffeomorphism with inverse
\be
\psi^{-1}(\mathbf{h})\,=\,\mathrm{e}^{\mathbf{h}}\,.
\ee
Following what we have done for the Wigner-Yanase metric tensor, we can use $\psi$ to transport every group action on $\stav$ to a group action on $\pos_{+}$.
In particular, $\stav$ is a real Euclidean space with respect to the (restriction of) the Hilbert-Schmidt product
\be
\langle\mathbf{h},\mathbf{k}\rangle\,=\,\mathrm{Tr}(\mathbf{h}\,\mathbf{k}).
\ee
Therefore, the Euclidean group acts on $\stav$ as
\be
A_{R,\mathbf{a}}(\mathbf{h})\,=\,R(\mathbf{h}) + \mathbf{a},
\ee
where $R$ is an element of the orthogonal group and $\mathbf{a}\in\stav$.

Quite interestingly, the unitary group may be realized as a subgroup of the orthogonal group of $\stav$ according to 
\be\label{eqn: coadjoint action of unitary group}
R_{\mathbf{U}}(\mathbf{h})\,:=\,\mathbf{U}\mathbf{h}\mathbf{U}^{\dagger}.
\ee
Indeed, it is easy to check that $R_{\mathbf{U}}$ preserves the Euclidean product
\be
\langle R_{U}(\mathbf{h}),R_{U}(\mathbf{k})\rangle\,=\,\langle\mathbf{U}\mathbf{h}\mathbf{U}^{\dagger},\mathbf{U}\mathbf{k}\mathbf{U}^{\dagger}\rangle\,=\,\langle\mathbf{h},\mathbf{k}\rangle.
\ee 
Consequently,  the group $\Uh\rtimes_{R} \stav$ acts on $\pos_{+}$ according to
\be
\widetilde{\Xi}((\mathbf{U},\mathbf{a}),\rho):=\varphi^{-1}\,\circ\,A_{R_{\mathbf{U}},\mathbf{a}}\,\circ\varphi(\rho)=\mathrm{e}^{\mathbf{U}\ln(\rho)\mathbf{U}^{\dagger} + \mathbf{a}}
\ee
Now, if we consider $\mathbf{a}=\mathbf{0}$,  we have
\be
\begin{split}
\Xi((\mathbf{U},\mathbf{0}),\rho)&=\mathrm{e}^{\mathbf{U}\ln(\rho)\mathbf{U}^{\dagger}} =\sum_{k=0}^{\infty}\frac{\left(\mathbf{U}\ln(\rho)\mathbf{U}^{\dagger}\right)^{k}}{k!}=\mathbf{U}\,\rho\,\mathbf{U}^{\dagger} = \phi(\mathbf{U},\rho),
\end{split}
\ee
and thus we obtain again the standard action $\phi$ of the unitary group on $\pos_{+}$.
The fundamental vector fields $\widetilde{\Upsilon}_{ab}$ of $\widetilde{\Xi}$ decompose as
\be
\widetilde{\Upsilon}_{\mathbf{ab}}\,=\,\mathcal{Z}_{\mathbf{a}} + \mathbb{X}_{\mathbf{b}}
\ee
where the $\mathbb{X}_{\mathbf{b}}$'s are the vector fields generating the standard action of the unitary group, and  the $\mathcal{Z}_{\mathbf{a}}$'s are given by 
\be\label{eqn: Z vector fields}
\mathcal{Z}_{\mathbf{a}}(\rho)\,=\,\frac{\mathrm{d}}{\mathrm{d}t}\left(\mathrm{e}^{ \ln(\rho) + t\mathbf{a}}\right)_{t=0}= \int_{0}^{1}\,\mathrm{d}\lambda\,\left(\rho^{\lambda}\,\mathbf{a} \,\rho^{1-\lambda}\right)\,,
\ee
where we used the well-known equality \cite{Suzuki-1997}
\be\label{eqn: derivative of exponential of operator}
\frac{\mathrm{d}}{\mathrm{d}t}\, \mathrm{e}^{A(t)} \,=\,\int_{0}^{1}\,\mathrm{d}\lambda\,\left(\mathrm{e}^{\lambda\,A(t)}\,\frac{\mathrm{d}}{\mathrm{d}t}(A(t))\,\mathrm{e}^{(1-\lambda)A(t)}\right)\,,
\ee
which is valid for every smooth curve $A(t)$ inside $\bh$ (remember that the canonical immersion of $\pos_{+}$ inside $\bh$ is smooth).

\begin{remark}
The Lie group  $\Uh\rtimes_{R} \stav$  is diffeomorphic to the cotangent bundle of the unitary group.
Indeed, if $G$ is any Lie group,the cotangent space $T^{*}G\cong G\times\mathfrak{g}^{*}$  is endowed with the structure of  Lie group \cite{A-G-M-M-1994,A-G-M-M-1998} according to
\be
(\gr_{1},\,a_{1})\cdot(\gr_{2},\,a_{2})\,:=\,(\gr_{1}\gr_{2},\,Ad_{\gr_{1} }^{*}(a_{2}) + a_{1}),
\ee
where $Ad^{*}$ is the dual of the adjoint action of $G$ on its Lie algebra $\mathfrak{g}$.
The resulting Lie group is also denoted by $G\rtimes_{Ad^{*}}\mathfrak{g}^{*}$  to emphasize the fact that the group structure is associated with a semidirect product.
Now, when $G=\Uh$, its Lie algebra $\mathfrak{g}$ is given by skew-adjoint operators on $\hh$ according to
\be
i\mathbf{b}\,\mapsto\,\mathbf{U}=\mathrm{e}^{i\mathbf{b}},
\ee
where $\mathbf{b}$ is an Hermitian operator.
Then, we can identify the dual space $\mathfrak{g}^{*}$ with the vector space $\stav$ of Hermitian operators by means of  the pairing
\be
\langle\mathbf{a},i\mathbf{b}\rangle\,:=\,\mathrm{Tr}(\mathbf{a}\,\mathbf{b}).
\ee
Consequently, the coadjoint action reads
\be
Ad_{\mathbf{U} }^{*}(\mathbf{a})\,=\,\mathbf{U}\,\mathbf{a}\,\mathbf{U}^{\dagger}\,=\, R_{\mathbf{U}}(\mathbf{a}),
\ee
where we used  equation \eqref{eqn: coadjoint action of unitary group} in the last equality, and  we conclude that $\Uh\rtimes_{R} \stav$  is actually diffeomorphic to the Lie group $T^{*}\Uh\equiv\Uh\rtimes_{Ad^{*}}\mathfrak{g}^{*}$ as claimed.

\end{remark}

We will now prove that every $\mathcal{Z}_{\mathbf{a}}$ is the gradient vector field associated with the smooth function $f_{\mathbf{a}}$ by means of the Bogoliubov-Kubo-Mori metric tensor $\G_{BKM}$ \cite{F-M-A-2019,Naudts-2018,N-V-W-1975,Petz-1993,P-T-1993}.
The  Bogoliubov-Kubo-Mori metric tensor  $\G_{BKM}$  is the metric tensor associated with the von Neumann-Umegaki relative entropy \cite{Araki-1976,Umegaki-1962-4,von-Neumann-1955}
\be\label{eqn: vN-U relative entropy}
S_{vNU}(\rho,\,\sigma)\,=\, \mathrm{Tr}\left(\rho\ln\rho - \rho\ln\sigma\right)
\ee

Just as we did for the Bures-Helstrom metric tensor and for the Wigner-Yanase metric tensor, we have to show that  
\be
\G_{BKM }\left(X,\,\mathcal{Z}_{\mathbf{a}}\right)\,=\,\mathcal{L}_{X}f_{\mathbf{a}},
\ee
where $X$ is an arbitrary vector field and $\mathcal{L}_{X}$ denote the Lie derivative with respect to $X$.
Therefore, we  compute
\be
\begin{split}
\left(\G_{BKM}\left(X,\mathcal{Z}_{\mathbf{a}}\right)\right)(\rho)  & = -\frac{\mathrm{d}}{\mathrm{d}t}\frac{\mathrm{d}}{\mathrm{d}s}\left(S_{vNU}(\gamma_{\rho}^{X}(t),\gamma_{\rho}^{\mathcal{Z}_{\mathbf{a}}}(s))\right)_{t,s=0} \\
& = \frac{\mathrm{d}}{\mathrm{d}t}\frac{\mathrm{d}}{\mathrm{d}s}\left(   \mathrm{Tr}\gamma_{\rho}^{X}(t)  \left(\ln(\rho) +s  \mathbf{a}\right) \right)_{t,s=0} \\
&\,=\,\mathrm{Tr}\left(X(\rho)\,\mathbf{a}\right),
\end{split}
\ee 
where we used the definition of $S_{vNU}$ given in equation \eqref{eqn: vN-U relative entropy},  and the equality
\be
\gamma_{\rho}^{\mathcal{Z}_{\mathbf{a}}}(s)=\mathrm{e}^{ \ln(\rho) +s  \mathbf{a}}
\ee
stemming from equation \eqref{eqn: Z vector fields}.
Recalling equation \eqref{eqn: lie derivative of qrv}, we conclude that 
\be
\G_{BKM }\left(X,\,\mathcal{Z}_{\mathbf{a}}\right)\,=\,\mathcal{L}_{X}f_{\mathbf{a}}
\ee
holds for all vector fields $X$ as desired.

\subsection*{Normalized states}

Once again, the normalized case follows from the not-normalized one.
Indeed, we just need to define the normalized action $\Xi$ of $T^{*}\Uh$ on $\stsp_{+}$ given by
\be
\Xi((\mathbf{U},\mathbf{a}),\rho)\,:=\,\frac{\widetilde{\Xi}(\gr,\rho)}{\Tr(\widetilde{\Xi}(\gr,\rho))}\,=\,\frac{\mathrm{e}^{\mathbf{U}\ln(\rho)\mathbf{U}^{\dagger} + \mathbf{a}}}{\Tr(\mathrm{e}^{\mathbf{U}\ln(\rho)\mathbf{U}^{\dagger} + \mathbf{a}})}.
\ee
Then, the fundamental vector fields  $\Upsilon_{\mathbf{ab}} $ of $\Xi $ are easily computed to be 
\be
\Upsilon_{\mathbf{ab}} \,=\,\mathbb{Z}_{\mathbf{a}}  + \mathbb{X}_{\mathbf{b}}
\ee
where the $\mathbb{X}_{\mathbf{b}}$'s are the vector fields generating the standard action of the unitary group, and the $\mathbb{Z}_{\mathbf{a}} $'s are given by
\be 
\mathbb{Z}_{\mathbf{a}} (\rho)\,=\,\frac{\mathrm{d}}{\mathrm{d}t}\left(\frac{\mathrm{e}^{ \ln(\rho)  + t\mathbf{a}}}{\Tr\left(\mathrm{e}^{ \ln(\rho)  + t\mathbf{a}}\right)}\right)_{t=0}=\int_{0}^{1}\,\mathrm{d}\lambda\,\left(\rho^{\lambda}\,\mathbf{a} \,\rho^{1-\lambda}\right) - \Tr(\rho\,\mathbf{a})\,\rho\,.
\ee

\begin{remark}
It is worth mentioning the recent work \cite{A-L-2020} where the finite transformations associated with the vector fields $\mathbb{Z}_{\mathbf{a}}$ are exploited in the definition of a Hilbert space structure on $\stsp_{+}$.
\end{remark}

Then, the Bogoliubov-Kubo-Mori metric tensor on $\stsp_{+}$ is associated with the pullback  to $\stsp_{+}$ (with respect to the canonical immersion) of the von Neumann-Umegaki relative entropy given in equation \eqref{eqn: vN-U relative entropy}.
With an evident abuse of notation, we denote by $S_{vNU}$  the pullback to $\stsp_{+}$ of the   von Neumann-Umegaki relative entropy given  by
\be
S_{vNU}(\rho,\,\sigma)\,=\, \mathrm{Tr}\left(\rho\ln\rho - \rho\ln\sigma\right) 
\ee
and by $\G_{BKM}$ the associated metric tensor on $\stsp_{+}$.

Once again, the computations performed in the not-normalized case may be easly adapted to prove that
\be
\G_{BKM} (X,\mathbb{Z}_{\mathbf{a}} )\,=\,\mathcal{L}_{X}f_{\mathbf{a}} \,,
\ee
for every vector field $X$ on $\stsp_{+}$, and where, with an evident abuse of notation, $f_{\mathbf{a}} $ is the pullback to $\stsp_{+}$ of the smooth function $f_{\mathbf{a}}$ on $\pos_{+}$.
From this, we conclude that every vector field $\mathbb{Z}_{\mathbf{a}} $ is the gradient vector field associated with the smooth function $f_{\mathbf{a}} $ by means of the Bogoliubov-Kubo-Mori metric tensor $\G_{BKM}$.

\section{Conclusions}\label{sec: conclusions}

The results of this work should be interpreted as a preliminary step toward a more general analysis aimed at characterizing those monotone metric tensors on the manifold of faithful quantum states that are linked with group actions in the sense explained in the previous sections.
Indeed, the fact that three of the most used metric tensors in Quantum Information Geometry like the Bures-Helstrom metric tensor, the Wigner-Yanase metric tensor, and the Bogoliubov-Kubo-Mori metric tensor are related with  group actions seems to point at a more profound connection which is yet to be discovered.
However, it is not yet known if this link is just a mathematical curiosity, or if it has to do with some  geometrical aspects of the monotonicity property usually required in Quantum Information Theory.

At this purpose, it is not hard to see that the requirement of unitary invariance encoded in the monotonicity property has a direct effect on the possible group actions considered as will be now explained.
It is clear that the Lie algebra of  the suitable extension  of $\Uh$ we can hope to link to a given monotone metric tensor $\G$ must be isomorphic, as a vector space, to $\uh\times\stav$.
Indeed, besides the fundamental vector fields $\mathbb{X}_{\mathbf{b}}$ of the action of $\Uh$  and associated with elements in $\uh$, we have  the {\itshape complementary} vector fields that must be the gradient vector fields associated with the expectation value functions $f_{\mathbf{a}}$ by means of $\G$, and the latters are clearly labelled by elements in the space $\stav$ of Hermitian operators on $\hh$.
Let us write these vector fields generically as $\mathbb{V}_{\mathbf{a}}$.
By assumption, we must have
\be
\G(X,\mathbb{V}_{\mathbf{a}})=\mathcal{L}_{X}f_{\mathbf{a}}
\ee
for every vector field $X$ on $\stsp_{+}$.
Then, we evaluate the Lie derivative of  the function $f_{\mathbf{c}}$ with respect to the commutator   $[\mathbb{X}_{\mathbf{b}},\mathbb{V}_{\mathbf{a}}]$ to obtain
\be
\begin{split}
\mathcal{L}_{[\mathbb{X}_{\mathbf{b}},\mathbb{V}_{\mathbf{a}}]}f_{\mathbf{c}}& \,=\,\mathcal{L}_{\mathbb{X}_{\mathbf{b}}}\left(\mathcal{L}_{\mathbb{V}_{\mathbf{a}}}f_{\mathbf{c}}\right) - \mathcal{L}_{\mathbb{V}_{\mathbf{a}}}\left(\mathcal{L}_{\mathbb{X}_{\mathbf{b}}}f_{\mathbf{c}}\right) \\
&=\mathcal{L}_{\mathbb{X}_{\mathbf{b}}}\left(\G(\mathbb{V}_{\mathbf{a}},\mathbb{V}_{\mathbf{c}})\right) - \mathcal{L}_{\mathbb{V}_{\mathbf{a}}}\left(f_{\frac{i}{2}[\mathbf{b},\mathbf{c}] }\right)\\
&= \mathcal{L}_{\mathbb{X}_{\mathbf{b}}}\left(\G(\mathbb{V}_{\mathbf{a}},\mathbb{V}_{\mathbf{c}})\right)  - \G(\mathbb{V}_{\mathbf{a}},\mathbb{V}_{\frac{i}{2}[\mathbf{b},\mathbf{c}]}),
\end{split}
\ee
where, in the second equality, we used the fact that
\be
\begin{split}
\left(\mathcal{L}_{\mathbb{X}_{\mathbf{b}}}f_{\mathbf{c}}\right)(\rho)&\,=\,\frac{\mathrm{d}}{\mathrm{d}t}\,\left(\Tr(\mathrm{e}^{i\frac{t}{2}\mathbf{b}}\,\rho\,\mathrm{e}^{-i\frac{t}{2}\mathbf{b}}\,\mathbf{c})\right)_{t=0} \\
& =\frac{i}{2}\Tr([\mathbf{b},\rho],\mathbf{c})  = \frac{i}{2}\Tr(\rho[\mathbf{b},\mathbf{c}]) =f_{\frac{i}{2}[\mathbf{b},\mathbf{c}] }(\rho)\,.
\end{split}
\ee
The unitary invariance of $\G$ implies that $\mathcal{L}_{\mathbb{X}_{\mathbf{b}}}\G=0$, so that 
\be
\begin{split}
\mathcal{L}_{[\mathbb{X}_{\mathbf{b}},\mathbb{V}_{\mathbf{a}}]}f_{\mathbf{c}}& \,=\,\G([\mathbb{X}_{\mathbf{b}},\mathbb{V}_{\mathbf{a}}],\mathbb{V}_{\mathbf{c}})  + \G(\mathbb{V}_{\mathbf{a}},[\mathbb{X}_{\mathbf{b}},\mathbb{V}_{\mathbf{c}}])  - \G(\mathbb{V}_{\mathbf{a}},\mathbb{V}_{\frac{i}{2}[\mathbf{b},\mathbf{c}]}) \\
& = \mathcal{L}_{[\mathbb{X}_{\mathbf{b}},\mathbb{V}_{\mathbf{a}}]}f_{\mathbf{c}}  + \G(\mathbb{V}_{\mathbf{a}},[\mathbb{X}_{\mathbf{b}},\mathbb{V}_{\mathbf{c}}])  - \G(\mathbb{V}_{\mathbf{a}},\mathbb{V}_{\frac{i}{2}[\mathbf{b},\mathbf{c}]}),
\end{split}
\ee
from which we conclude that
\be
\G(\mathbb{V}_{\mathbf{a}},[\mathbb{X}_{\mathbf{b}},\mathbb{V}_{\mathbf{c}}]  -  \mathbb{V}_{\frac{i}{2}[\mathbf{b},\mathbf{c}]})\,=\,0\,.
\ee
Since the differentials of the $f_{\mathbf{a}}$ form a basis of the module of one-forms on $\stsp_{+}$,  and  since the $\mathbb{V}_{\mathbf{a}}$ are the gradient vector fields associated with the $f_{\mathbf{a}}$, we have that the $\mathbb{V}_{\mathbf{a}}$ form a basis of the module of vector fields on $\stsp_{+}$, and thus the previous equation implies
\be
[\mathbb{X}_{\mathbf{b}},\mathbb{V}_{\mathbf{c}}] =  \mathbb{V}_{\frac{i}{2}[\mathbf{b},\mathbf{c}]}.
\ee
This means that the commutator between the $\mathbb{X}_{\mathbf{b}}$ and the {\itshape complementary} vector fields is fixed by the requirement of unitary invariance.
Therefore, the Lie algebra structure on $\uh\times \stav$ must fulfill this additional constraint.
Taking into account that $\uh$ must be realized as a Lie subalgebra of $\uh\times\stav$, we conclude that the freedom we have in choosing the enlargment of $\uh$ is only in the Lie product between  elements in the complementary space $\stav$.
In particular, for the Lie algebra of $\Glh$ considered in section \ref {sec: B-H metric tensor} and section \ref{sec: W-Y metric tensor}, this bracket gives back an element in $\uh$, while, for the Lie algebra of $T^{*}\Uh$  considered in section \ref{sec: B-K-M metric tensor}, this bracket vanishes identically.
A classification of the possible extensions of the Lie algebra $\uh$ satisfying the constraints found above will certainly give a hint on what type of group actions one can hope to link to monotone metric tensors.
According to \cite{A-G-M-M-1994,A-G-M-M-1998}, the groups $\Glh$ and $T^{*}\Uh$ are symplectomorphic and Morita-equivalent, and each of them is a so-called group double of $\Uh$.
This leads to conjecture that the appropriate enlargments may be looked for in the context of the double metric-Lie algebras containing $\uh$ as a subalgebra.
The emergence of a group double of $\Uh$ in both situations we discussed, seems to suggest that we may deal with the Non-Commutative Extension of Information Geometry by using the whole machinery introduced by Drinfeld to deal with Poisson-Lie groups (see again \cite{A-G-M-M-1994,A-G-M-M-1998} and references therein).

From another point of view,  the case of monotone metric tensors obtained from the quantum Tsallis relative entropies \cite{M-M-V-V-2017} and from the $\alpha-z$-relative R\'{e}ny relative entropies \cite{A-D-2015,C-DC-L-M-M-V-V-2018} is  currently being investigated using the  methodology developed in section \ref{sec: W-Y metric tensor}.


Finally, it should be noted that it is possible to reformulate the results contained in this work  no more in terms of quantum states on $\hh$, but in terms of the more general notion of states on an arbitrary  finite-dimensional  von Neumann algebra.
This would help to give a unifying picture for the classical and quantum case, and would be necessary in order to extend this information geometrical considerations also to the recent groupoidal approach to quantum theories developed in \cite{C-DC-I-M-2020,C-DC-I-M-02-2020,C-I-M-2018,C-I-M-02-2019,C-I-M-03-2019,C-I-M-05-2019}.

%

%

\end{document}